\documentclass[%
 preprint, 
superscriptaddress,
 amsmath,amssymb,
 aps, physrev,
]{revtex4-2}
\usepackage{float}
\usepackage{todonotes}
\usepackage{soul,color}
\usepackage{dcolumn}
\usepackage{bm}
\usepackage{comment}
\usepackage{graphicx}
\usepackage{caption}
\usepackage{subfig}
\usepackage{subcaption}
\usepackage{hyperref}

\usepackage{bbold}

\usepackage{caption}
\usepackage{subcaption}
\usepackage{amsfonts,amsmath,amssymb,cancel}
\DeclareMathOperator{\sech}{sech}

\begin{document}

\preprint{APS/123-QED}


\title{Schr\"odinger equation is $\mathcal{R}$-separable in toroidal coordinates}

\author{Matheus E. Pereira}
 \email{mepereira@id.uff.br}
\author{Alexandre G. M. Schmidt}%
\affiliation{%
  Instituto de Ci\^encias Exatas, Universidade Federal Fluminense,
  \\ 27213-145 Volta Redonda --- RJ, Brazil
}%
\affiliation{Programa de P\'os Gradua\c{c}\~ao em F\'isica, Instituto de F\'\i sica, Universidade Federal Fluminense,\\
	24210-346 Nitew\'oi --- RJ, Brazil}




\date{\today}
\begin{abstract}
We present, for the first time, exact solutions for the Schr\"{o}dinger equation in Moon and Spencer's toroidal coordinates, and in the electromagnetic toroidal--poloidal coordinate systems. Curiously, both systems present a fractional angular momentum, because of the torus's hole. We achieve these novel solutions using the irregular $\mathcal{R}$-separation of variables, an unexplored approach in Physics, which results in a wavefunction with fractional angular momentum eigenvalues. Numerous solutions for the Schr\"{o}dinger equation in a variety of external potentials are shown, including an external magnetic field. A plane-wave expansion and a Green function are also presented, setting the stage for future progress in this area.
\end{abstract}

\maketitle



\section{Introduction}\label{sec:introduction}

Usually, one of the steps to solve Schr\"{o}dinger equation using simple separation of variables is to consult Moon and Spencer's handbook \cite{moon2012field} and find if the equation is said to be separable in a given coordinate system. Historically, it has been shown \cite{MOON1952-2,moonspencer_06_1952,moonspencer_09_1952, eisenhart1,eisenhart2} that the free Schr\"{o}dinger (Helmholtz) equation is simply separable in exactly eleven coordinate systems, while $\mathcal{R}$-separation is possible only for rotational coordinate systems, like spherical and spheroidal. Therefore, the challenge to solve the Helmholtz equation in any coordinate system that defies Moon and Spencer's impossibility theorems remained open for decades. While most textbook problems revolve around symmetric configurations, which are contemplated by one of those eleven coordinate systems \cite{arscott-darai}, real-life physical applications are seldom trivial \cite{nanotorus,jacobi-dandoloff,li2024twisted}. In particular, we found important contributions in the solution of the Laplace and hydrodynamic equations in bispherical coordinates \cite{thermal,stoy-bispherical,MAJIC-bispherical,bispherical-transport}. The theory of regular and irregular $\mathcal{R}$-separation can be found in the seminal works of Kalnins and Miller \cite{kalnins1976lie,kalnins1976symmetry,kalnins1980killing,kalnins1980some,kalnins1984theory}.

The toroidal geometry is widely known, for example, for shaping both donuts and tokamaks \cite{tokamak1,tokamak2}, which are devices designed to manipulate plasma at extremely high temperatures (hundreds of thousands of kelvins) \cite{temp1,temp2}. Because of its deceptively simple and symmetric shape, attempts to solve Helmholtz equation using Moon and Spencer's toroidal coordinates or the electromagnetic toroidal--poloidal coordinates via simple separation of variables only resulted in failure. 

In the previous two decades, there have been several attempts to investigate the Schr\"odinger equation in toroidal coordinates, either on the torus surface or in a toroidal quantum well. Encinosa and collaborators studied numerically \cite{encinosa-PRA1} bound states induced by curvature and in the presence of a magnetic field \cite{encinosa-PRA2}. Schulze-Halberg \cite{schulze1, schulze2} also tackled the problem of a quantum particle adhered to a toroidal surface; however, the azimuthal quantum number was taken to be zero, and the geometrical potential was not taken into account. Encinosa also studied the quantum mechanical problem in an elliptical torus \cite{encinosa-phys-scr1} and the role of hard and soft boundary conditions \cite{encinosa-phys-scr2}, both using a numerical approach. Atanasov and collaborators also made an important contribution where they considered the geometrical potential, however, their results are valid for weak magnetic fields and certain approximations in the geometrical parameters \cite{atanasov}. In our paper we do not resort to approximations in these geometrical parameters, nor in particular choices of quantum numbers. 
The next section reveals the only known closed-form solution to the Helmholtz equation in Moon and Spencer's toroidal coordinate system. Additionally, we present an alternative coordinate system in the same section to serve as a basis to the ensuing discussions. Solutions for a free particle and for a particle in an infinite potential well are shown in section \ref{sec:free}. The first is concisely written in terms of Bessel and exponential functions. The last are written using Heun functions \cite{ronveaux}. Section \ref{sec:funcao-de-green} presents a Green function and the plane-wave expansion. We conclude with a discussion and our final remarks in section \ref{sec:conclusao}.

\section{Preliminary considerations and an alternative toroidal coordinates}\label{sec:preliminar}






\subsection{A toroidal trouble}
Consider Moon and Spencer's toroidal coordinate system
\begin{align*}
x &= \frac{R \sinh \tau \cos \phi}{\cosh \tau - \cos \theta}, \\
y &= \frac{R \sinh \tau \sin \phi}{\cosh \tau - \cos \theta}, \\
z &= \frac{R \sin \theta}{\cosh \tau - \cos \theta}.
\end{align*}
Here, $R$ is a constant that describes the distance from the cartesian origin to a ring of radius $R$ at the $xy$ plane --- the focal ring. More details can be found in the works of Moon and Spencer and of Weston \cite{moonspencer_06_1952,moonspencer_09_1952,MOON1952-2,weston1958-1,weston1958-2, weston1960-on-toroidal}. Now consider the Helmholtz equation in these coordinates \cite{moon2012field},
\begin{multline}\label{equacao-de-helmholtz-toroidal-moon}
    \frac{(\cosh \tau - \cos \theta)^3}{R^2 \sinh \tau} \Bigg[
\sinh \tau \, \frac{\partial}{\partial \theta} \left( \frac{1}{\cosh \tau - \cos \theta} \frac{\partial \psi}{\partial \theta} \right) 
+ \frac{\partial}{\partial \tau} \left( \frac{\sinh \tau}{\cosh \tau - \cos \theta} \frac{\partial \psi}{\partial \tau} \right) \\
+ \frac{1}{\sinh \tau (\cosh \tau - \cos \theta)} \frac{\partial^2 \psi}{\partial \phi^2}
\Bigg] + k^2 \psi = 0.
\end{multline}
Ever since Moon and Spencer laid down their well-known works on separation of variables, it has been believed that a separable solution was impossible to achieve for such an equation. This is still true. However, more recent investigations by Kalnins, Miller, Prus and Sym \cite{prus-e-sym} introduced a new type of solutions for partial differential equations, known as an irregular $\mathcal{R}$-separation of variables. Following the results of Prus and Sym, calculations revealed
\begin{equation}
\begin{aligned}
    \psi(\tau,\theta,\phi) &= N_\pm  \exp{\left(\pm i  k R \frac{\sinh \tau }{\cosh \tau - \cos \theta}\right)} \sqrt{e^{i\phi}\frac{\cosh \tau - \cos \theta}{\sinh \tau }}\\
    &= e^{i\phi/2}\left[c_1 J_{1/2}\left(kR\frac{\sinh \tau}{\cosh \tau - \cos \theta}\right) + c_2 H_{1/2}^{(1)}\left(kR\frac{\sinh \tau}{\cosh \tau - \cos \theta}\right)\right]
    \end{aligned}\label{solucao-matheus}
\end{equation}
is the solution of Helmholtz equation \eqref{equacao-de-helmholtz-toroidal-moon} in toroidal coordinates using irregular $\mathcal{R}$-separation of variables \cite{prus-e-sym}, where $J_\nu$ is the ordinary Bessel function of order $\nu$ and $H_\nu^{(1)}$ is the Hankel function of the first kind with order $\nu$. A simple investigation reveals a troubling fact though, that this solution cannot be used for solving physical problems, in agreement with \cite{prus-e-sym}. Indeed, consider the problem of a particle in an infinite toroidal well of minor radius $a$ and major radius $R$. Because $(0,0,0)$ is actually outside the well, we have $c_2$ may not be zero. For integer $n$, we get
\begin{equation}
    \psi(a,\theta,\phi) = 0 \quad \rightarrow \quad kR\frac{\sinh a}{\cosh a - \cos \theta} = n\pi \quad \rightarrow\quad \cos\theta = \cosh a - \left(\frac{kR}{n\pi}\right)\sinh a,
\end{equation}
revealing the well does not trap the particle except where $\cos\theta$ obeys the last expression, which is not a valid condition for the particle in a well problem. 

\subsection{The toroidal--poloidal coordinate system}

\begin{figure}[h]
    \centering
    \includegraphics[width=0.5\textwidth]{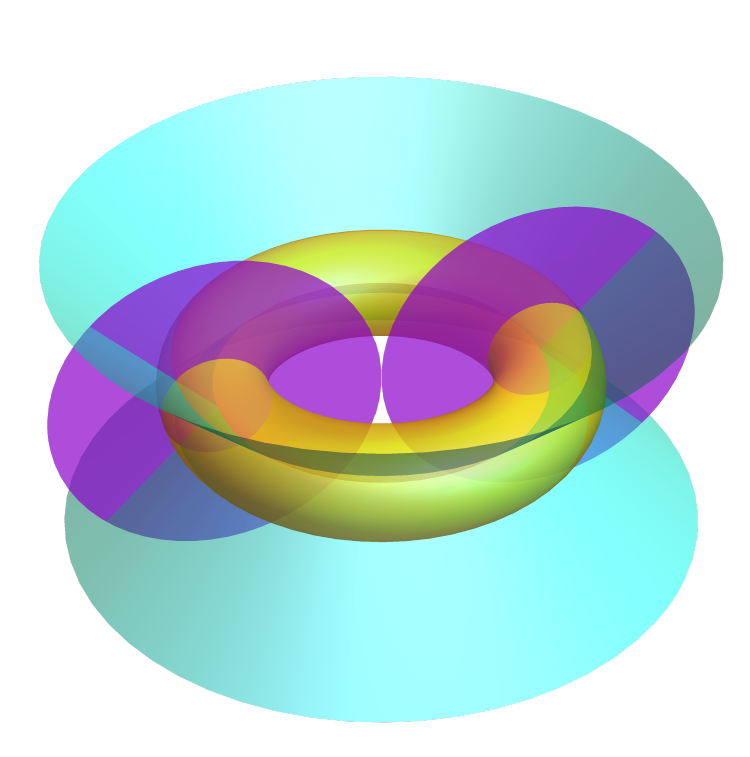}
    \caption{Level surfaces of the toroidal/poloidal coordinate system. We see a yellow torus, as a surface of constant $w$, cyan cones, of constant $u$, and magenta disks, setting $v$ constant.}
    \label{fig:superficies}
\end{figure}

\begin{figure}[h]
    \centering
    \includegraphics[width=1\linewidth]{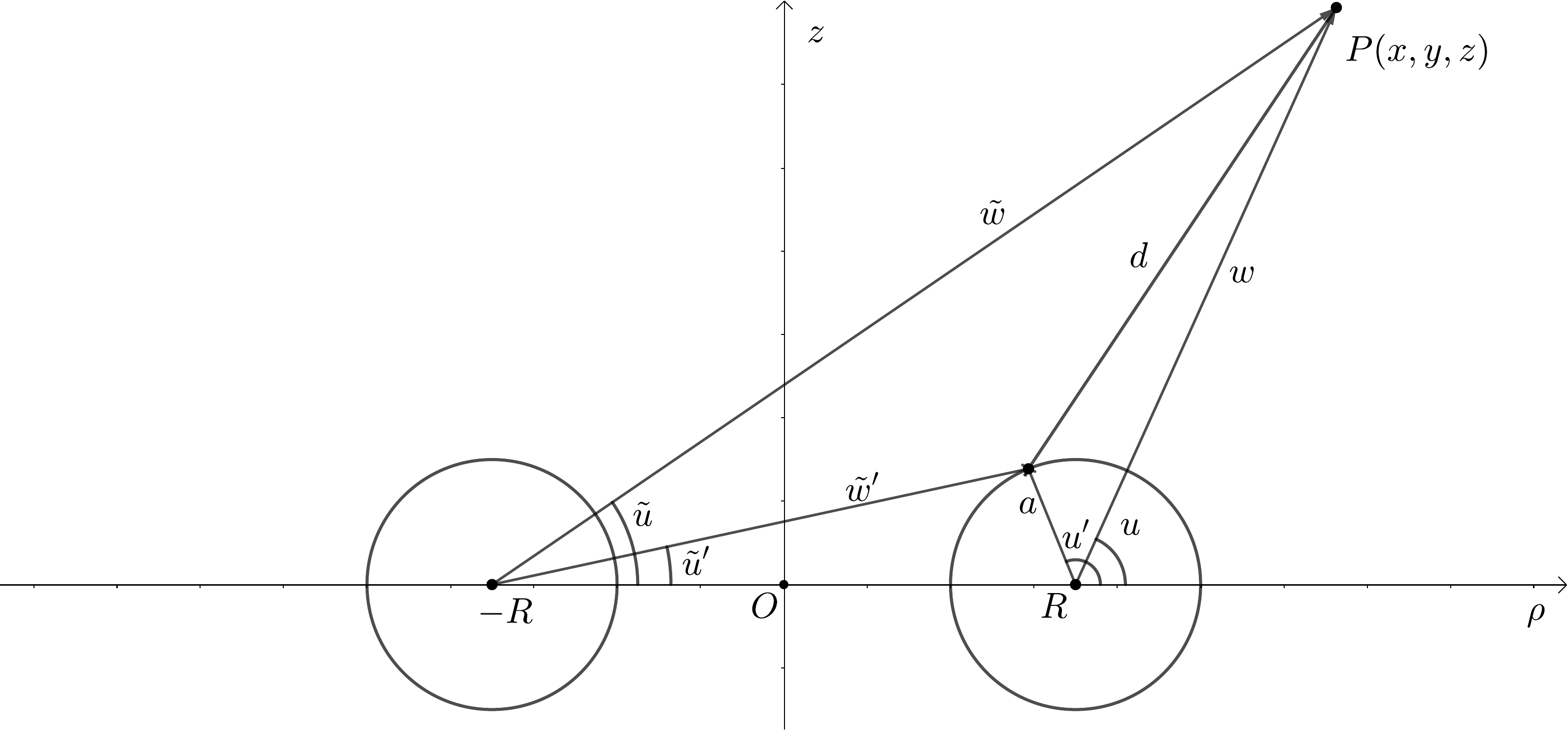}
    \caption{Cross section of a torus with minor radius $a$, major radius $R$ for a fixed $v$, where the plane $\rho$ is defined by $z = 0$ and $\rho = \sqrt{x^2 + y^2}$.}
    \label{fig:secao-transversal}
\end{figure}

To remedy the situation explained in the previous section, consider the Dupin cyclide parametric equations,
\begin{align*}
x &= \frac{b^2 \cos v \cosh \eta + w (c \cosh \eta - a \cos v)}{a \cosh \eta - c \cos v}, \\
y &= s_y\frac{b \sin v (a \cosh \eta - w)}{a \cosh \eta - c \cos v}, \\
z &= s_z\frac{b \sinh \eta (w - c \cos v)}{a \cosh \eta - c \cos v}.
\end{align*}
where $s_z$ and $s_y$ are dimensionless scaling quantities and $c^2 = a^2 - b^2$. Prus and Sym \cite{prus-e-sym} studied the irregular $\mathcal{R}$-separation in this coordinate system, achieving some exact solutions for Helmholtz equation. We lift most of their restrictions, allowing the Dupin cyclide to reduce to a very special case. If $a = b = R$, these equations describe a ring circular torus whose minor and major radii are measured by $w$ and $R = b$ respectively. Namely,
\begin{align*}
x &= (R  - w\sech \eta)  \cos v, \\
y &= s_y(R  - w\sech \eta)  \sin v, \\
z &= s_z w\tanh \eta  .
\end{align*}
This is not a good choice of coordinates, because it lacks periodicity in $\eta$, which is a necessary feature of toroidal symmetry. Morever, Prus and Sym \cite{prus-e-sym} did solve the Helmholtz equation for the more general case, and their solution is not appropriate for physical problems, as they recognize. Therefore, we must make a dramatic change in these coordinates, in order to recover periodicity. Following Maxwell \cite{maxwell}, by transforming $\cosh \eta \rightarrow \sec u$ and $\sinh{\eta} \rightarrow \tan u$  we acquire
\begin{equation}\label{eq parametricas}
    \begin{aligned}
        x &= (R + w\cos u) \cos  v,\\
        y &= s_y(R + w\cos  u) \sin  v,\\
        z &= s_z w \sin  u,
    \end{aligned}
\end{equation}
where $s_z = \pm 1, s_y = \pm 1$ and we made $w \rightarrow -w$ for convenience. These are the parametric equations of a regular circular ring torus, which also happen to describe a right-handed orthogonal coordinate system if $s_z = 1$, $s_y = -1$. They define the so-called \textit{toroidal and poloidal coordinate system}, where $0 \leq w < \infty$, $0 \leq u, v \leq 2\pi$. Incidentally, they are an alternative to the usual toroidal coordinate system, and can also be obtained from the conformal mapping
\begin{equation}
    z = x + iy = e^w + R,
\end{equation}
where $w = u + iv$. This means that the toroidal-poloidal coordinates are conformal, thus angle-preserving. Using a position vector $\mathbf{x} = (x, y, z)$, the scale factors can be calculated from \eqref{eq parametricas} as
\begin{equation}\label{fatores-de-escala}
        h_w = \left|\frac{\partial \mathbf{x}}{\partial w}\right| = 1, \quad  h_u = \left|\frac{\partial \mathbf{x}}{\partial u}\right| = w, \quad  h_v = \left|\frac{\partial \mathbf{x}}{\partial v}\right| = (R + w\cos u).
\end{equation}
Consequently, the determinant of the metric tensor is
\begin{equation}\label{det-g}
    \sqrt{\det g} = \sqrt{g} = h_w h_u h_v = w(R + w\cos u).
\end{equation}
Thus, the volume element $dV = \sqrt{g}\; dw\; du\; dv $ and surface element $dS = \sqrt{g}|_{w = w'} \; du\; dv$ are given by,
\begin{equation}\label{elementos-de-inregracao}
    dV = w(R + w\cos u)\; dw\; du \; dv, \qquad dS =  w'(R + w'\cos u) du \; dv.
\end{equation}
To gain insight on this system, we notice that for constant $z$, $\rho = \sqrt{x^2 + y^2}$ defines circles in the $xy$ plane, where we define $\tan u = y/x$. We obtain cone-like surfaces if $u$ is a constant, which degenerate to disks for $u = 0$. On the other hand, surfaces of constant $v$ are all disks centered at $w = 0$. Finally, constant $w$ draws toroidal surfaces, as we see in Figure \ref{fig:superficies}. 
We choose $s_z = -1$, thus relating our coordinates to Cartesian coordinates via
\begin{align*}
    w &= \sqrt{(\rho - R)^2 + z^2}, \\
    u &= \arctan\left(\frac{-z}{\rho - R}\right), \\
    v &= \arctan\left(\frac{y}{x}\right).
\end{align*}
Consider a point $P$ in space like Figure \ref{fig:secao-transversal}. We notice that
\begin{gather}
\Tilde{w}(w,u) = \sqrt{w^2 + 4R^2 - 4Rw\cos u}, \label{funcao w til}\\
w(\Tilde{w},\Tilde{u}) = \sqrt{\Tilde{w}^2 + 4R^2 - 4R\Tilde{w}\cos\Tilde{u}}, \\
\text{and} \nonumber \\
\cos\Tilde{u} = \frac{2R + w\cos u}{\Tilde{w}(w,u)} . \label{angulo novo}
\end{gather}
Also, this figure shows a troubling fact about this coordinate system. Any point with Cartesian coordinates $P(x,y,z)$ has two representations in toroidal-poloidal coordinates. In other words, we are essentially using two coordinate systems at once by employing \eqref{eq parametricas}, which is well-known in the mathematics community as the \textit{double covering} of the torus.
 While from the point $(R\cos v, R\sin v,0)$ the distance to $P$ is $w$, from the point $-(R\cos v, R\sin v,0)$, which is its antipodal, located at $(w,u,v+\pi)$, this distance is $\Tilde{w}$. This reveals a curious characteristic of this system: even if our wavefunction is regular at a given point, there will be another wavefunction that will be evaluated at the same point, but with a different result.
\section{The free particle and the toroidal well}\label{sec:free}
\subsection{Free particle}
\label{sub-section-free-particle}
The time-independent Schr\"{o}dinger equation for a free particle can be written as a Helmholtz equation with $k^2 = 2m^* E/\hbar^2$, where $m^*$ is the mass, $E$ is the energy and $\hbar$ is the reduced Planck's constant. In three-dimensional orthogonal curvilinear coordinates, this is 
\begin{equation}
    \label{equacao de helmhotz em sistema de coordenada curvilineo ortogonal}
    \frac{1}{\sqrt{g}}\sum_{i = 1}^3\frac{\partial}{\partial x_i}\left(\frac{\sqrt{g}}{g_{ii}}\frac{\partial\varphi}{\partial x_i}\right) + \left[k^2 + V(x_1,x_2,x_3)\right]\varphi = 0.
\end{equation}

 Substituting the metric tensor and its determinant, we easily have
\begin{widetext}
\begin{multline}\label{edp-helmholtz-toroidal}
 \frac{\partial^2 \varphi}{\partial w^2} + \left(\frac{1}{w} + \frac{\cos u}{R + w\cos u}\right)\frac{\partial \varphi}{\partial w} + \frac{1}{w^2}\frac{\partial^2 \varphi}{\partial  u^2} - \frac{\sin u}{w(R + w\cos u)}\frac{\partial \varphi}{\partial  u} + \frac{1}{(R + w\cos u)^2}\frac{\partial^2 \varphi}{\partial  v^2} \\ + V(w,u,v)\varphi + k^2\varphi = 0.
\end{multline}
\end{widetext}
Considering $V(w,u,v) = 0$ for the free particle and applying a $F-$homotopic transformation, we introduce the \textit{ansatz}
\begin{equation}\label{psi_m}
    \varphi(w, u, v) = w^A (R + w\cos u)^B f(w) e^{i \alpha  u} e^{C v},
\end{equation}
where the constants $A, B$, $C$ and $\alpha$ have to be chosen such that equation \eqref{edp-helmholtz-toroidal} is (non-regularly and non-St\"{a}ckel) $\mathcal{R}$-separable \cite{prus-e-sym}. The periodicity condition in $u$ implies $\alpha = m$, where $m$ is an integer. Taking $B = -1/2$ reveals a Bessel equation for $f(w)$ and also determines $C$, since $C = \pm i|B|$. Thus,
\begin{equation}
    \frac{d^2 f}{dw^2} + \left(\frac{1 + 2A}{w}\right)\frac{d f}{dw} + \left(k^2 + \frac{A^2 - m^2}{w^2}\right)f(w) = 0.
\end{equation}

Solutions of this equation can be written in terms of Bessel and Hankel functions, namely, $f(w) = w^{-A} \left[ c_1 J_m(kw) + c_2 H^{(1)}_m(kw)\right]$. Thus, substituting into \eqref{psi_m}, we obtain the free solution of the Schr\"{o}dinger equation
\begin{equation}\label{toroidal-wavefunctions}
    \varphi_m(w, u, v) = A_m\frac{J_{m}(kw) e^{i m u}\exp{\left(-i v/2\right)}}{\sqrt{R + w\cos u}}  ,
\end{equation}
where we dropped the irregular solution, and $A_m$ is a constant. The wave number $k$ is continuous for the free particle, but otherwise must be calculated considering appropriate boundary conditions. It is most extraordinary to realize that the wavefunction \textit{is not $2\pi$ periodic in $v$}. Indeed,
\begin{equation}\label{fase}
     \varphi_m(w, u, v + 2\pi) = e^{-i\pi} \varphi_m(w, u, v) .
\end{equation}
The appearance of this geometrical phase is explained by the fact that the irregular $\mathcal{R}$-separability removes one quantum number, which is reflected in the $\mathcal{R}$-factor. Notice we can write
\begin{equation}
    \mathcal{R}(w,u,v) = \left[(R + w\cos u)e^{i v}\right]^{-1/2} = (x + i y)^{-1/2} .
\end{equation}
As we circle the Cartesian origin, the multi-valued nature of the wavefunction becomes evident: every time $x = 0$, the square root of an imaginary number brings a phase $e^{-i\pi/2}$. This results in the wavefunction jumping to its other branch once the rotation completes $2\pi$, leaving a phase $e^{-i\pi}$ behind. $\varphi$ returns to the original branch upon a $4\pi$ rotation. The reason is that since the wavefunction will be singular at $x + iy = 0$ and we use a sine angular function to enforce regularity at these points, \textit{we have effectively excised the $z$-axis} from the configuration space. Indeed, this remarkable change means that
\begin{equation*}
    \mathbb R^3\setminus\{(0,0,z)\mid z\in\mathbb R\}\simeq \mathbb R^2\setminus\{(0,0)\}\simeq S^1,
\end{equation*}
that is, the three-dimensional Euclidean space without the $z$ axis is homotopically equivalent \cite{nash-sen} to a plane without the origin, which is homotopically equivalent to a circle. Because now the configuration space is multiply connected, a new topology comes about, which implies the phase \cite{kowalski} in \eqref{fase} to be a \textit{topological, or geometrical phase}, much like the historical Aharonov--Bohm effect.
While our wavefunction \eqref{toroidal-wavefunctions} is suitable to understand the dynamics of a quantum particle trapped in a toroidal well, that is, restricting the radial-like domain, this is still a multivalued solution if one wishes to treat a free particle. As we mentioned, the coordinate system is not injective. 

\begin{figure}%
    \centering
    \subfloat[\centering]{{\includegraphics[width=7.5cm]{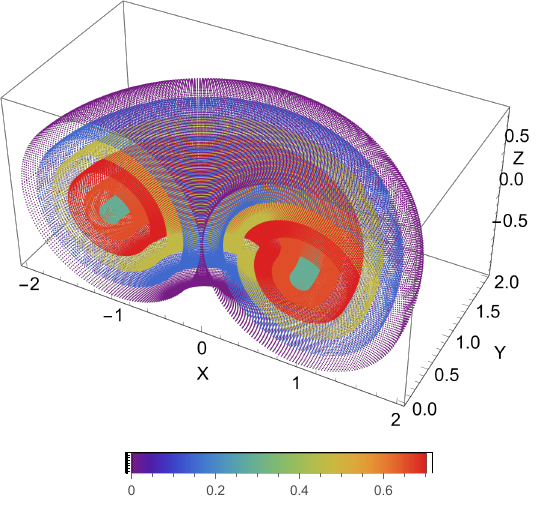} }}%
    \qquad
    \subfloat[\centering]{{\includegraphics[width=7.5cm]{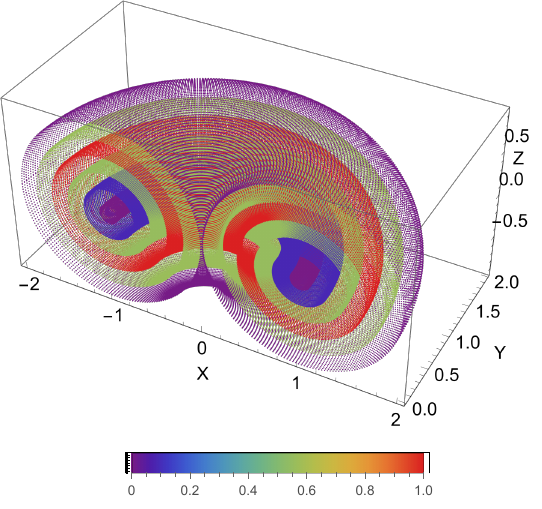} }}%
    \caption{Probability densities $\sigma |\psi_{0}(w,u,v)|^2$ and $\sigma |\psi_{2}(w,u,v)|^2$, $k_{0,1} \approx 2.4048$, the first zero of $J_0(x)$, where $\sigma=\sqrt{g}|_{w = w'}$ given by \eqref{det-g}. Observe that the probability densities vanish at the toroidal shell. Here we used $R=1, m^*=1/2$ and $\hbar=1$. }%
    \label{fig:1o grafico}%
\end{figure}

\begin{figure}%
    \centering
    \subfloat[\centering]{{\includegraphics[width=7.5cm]{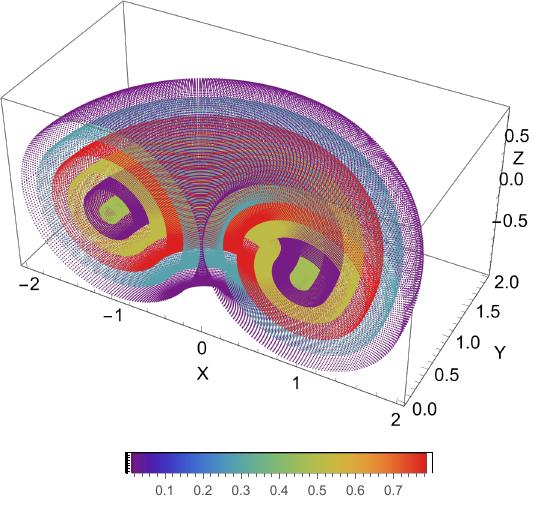} }}%
    \qquad
    \subfloat[\centering]{{\includegraphics[width=7.5cm]{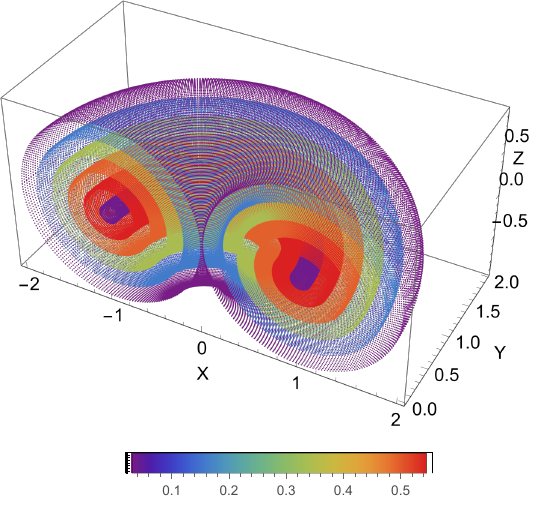} }}%
    \caption{Probability densities of $\sigma |\psi_{1}(w,u,v)|^2$ and $\sigma |\psi_{3}(w,u,v)|^2$, using $k = k_{4,0}$ and $k = k_{5,0}$, the fourth and fifth zeros of $J_0(x)$, where $\sigma$ is the surface element and we used the same values of the previous figure: $R=1, m^*=1/2$ and $\hbar=1$. We observe that there is a region where the probability densities vanish (purple tube) inside the well in both of them. Maxima takes place at red regions. }%
    \label{fig:2o grafico}%
\end{figure}

In order to gain insight about this problem, we plot four normalized probability densities calculated using (\ref{toroidal-wavefunctions}). Since the toroidal well is infinite, we impose that the wave function vanishes at its wall, namely $\varphi_m(R,u,v)=0$, i.e., the wave number must be a zero of the ordinary Bessel function. In figures \ref{fig:1o grafico} and \ref{fig:2o grafico} we restrict the $v$ angle to $[0,\pi]$ in order to see inside the well.

\subsection{Toroidal well with external potentials}
Toroidal geometries have been extensively studied for applications ranging from nuclear fusion (tokamaks) to nanodevices \cite{nanotorus,toroidalnano1,toroidalnano2, magnetohidrodinamica,Yoshida_estrela,eletrodinamica}. There even exists a toroidal black hole \cite{buraco-negro-toroidal}. However, as we mentioned, it was thought to be impossible to analytically solve the scalar wave equation using separation of variables, which is one of the most useful techniques for approaching the problem. Now, with the solution within our grasp, we show a variety of external potentials for which the Schr\"{o}dinger equation is still exactly solvable in toroidal-poloidal coordinates using this irregular ${\cal R}-$separation method.

\subsubsection{Case 1: $V_{ext} = V^{(1)}_{ext}(w,u)$}

Let us consider an external potential that depends on $w$ and $u$ as,
\begin{equation} \label{potencial-U1-nao-central}
    V^{(1)}_{ext}(w, u) = 4U_1w+ 4U_2 w^2+ \frac{V_1}{w} + \frac{V_2}{w^2} + \frac{T_2}{4(R+w\cos u)^2}, 
\end{equation}
where the parameters $U_1, U_2, V_1, V_2,$ and $T_2$ represent real coupling constants. Substitution of this potential into \eqref{equacao de helmhotz em sistema de coordenada curvilineo ortogonal} results in the non-normalized eigenfunctions
\begin{equation}\label{psi1_potencial-U1-nao-central}
    \psi^{(1)}_m(w, u, v) = c_m \exp{\left[ \frac{U_2(U_1+U_2w)w}{(-U_2)^{3/2}}  +im u - \frac{ iv\sqrt{1+T_2}}{2}\right]} \frac{w^{\sqrt{m^2-V_2}}\; {\rm HeunB}(\{ 1\};w)}{\sqrt{R+w\cos u}},
\end{equation}
where $\{1\}$ represents the set of five parameters of the bi-confluent Heun's function \cite{ronveaux, lay},
\begin{equation}
    \{1\} = \left\{q_1, k^2 -\frac{U_1^2}{U_2}  +4\sqrt{-U_2}\left(1+\sqrt{m^2-V_2}\right), 1 + 2\sqrt{m^2-V_2}, -\frac{2U_1}{\sqrt{-U_2}}, 4\sqrt{-U_2} \right\}.
\end{equation}
We define the accessory parameter as
\begin{equation}
    q_1 = -V_1 + \left(\frac{1 +2  \sqrt{m^2-V_2} }{\sqrt{-U_2}}\right)U_1. 
\end{equation}
The bi-confluent Heun's function $y(z)={\rm HeunB}(q, \alpha, \gamma, \delta, \varepsilon; z) $ satisfies the differential equation \cite{olver},
\begin{equation}
    zy^{''} + \left(\gamma+ \delta z+ \varepsilon z^2\right) y' + (\alpha z-q)y =0, 
\end{equation}
where $\gamma$ can not be zero nor a negative integer.

The second solution reads,
\begin{equation}\label{psi2_potencial-U1-nao-central}
    \psi^{(2)}_m(w, u, v) = d_m \exp{\left[\frac{U_2(U_1+U_2w)w}{(-U_2)^{3/2}}+im u - \frac{ iv\sqrt{1+T_2}}{2}\right]} \frac{w^{-\sqrt{m^2-V_2}}\; {\rm HeunB}(\{ 2\};w)}{\sqrt{R+w\cos u}},
\end{equation}
where $\{2\}$ represents another set of five parameters of the bi-confluent Heun's function \cite{ronveaux,lay},
\begin{equation}
    \{2\} = \left\{q_2, k^2+ 4\sqrt{-U_2}\left(1 -\sqrt{m^2-V_2}\right), 1 - 2\sqrt{m^2-V_2}, 0, 4\sqrt{-U_2} \right\}.
\end{equation}
where the accessory parameter is slightly changed,
\begin{equation}
    q_2 = -V_1 + \left(\frac{1 - 2\sqrt{m^2-V_2} }{\sqrt{-U_2}}\right)U_1.
\end{equation}

The bi-confluent Heun's equation has a regular singularity at $w=0$, and an irregular singularity at $\infty$ of rank $2$. It has five parameters. One of them, usually represented as $q$, the so-called accessory parameter. Its solutions are known as bi-confluent Heun's functions, which on their turn, are entire functions. In a previous paper \cite{black-string-PLA} we have shown how to obtain polynomial solutions of this equation. 


\subsubsection{Case 2: $V_{ext} = V^{(2)}_{ext}(w)$}

There is a second external potential we can introduce that allows exact solutions of the $w-$ODE. Consider the potential,
\begin{equation} \label{potencial-U2-inversos}
    V^{(2)}_{ext}(w) = V_0 + \frac{V_1}{w} + \frac{V_2}{w^2} + \frac{V_3}{w^3} + \frac{V_4}{4w^4}, 
\end{equation}
where $V_i$ are real constants. The ``radial-like" equation can be rewritten as a double-confluent Heun's equation, which is a second order ODE with two irregular singularities located at $w=0$ and $\infty$, each of rank 1. The first  non-normalized eigenfunction is,
\begin{equation}\label{psi1_potencial-U2-inversos}
    \psi^{(1)}_m(w, u, v) = c_m \exp{\left(imu - \frac{iv}{2} - i\kappa w +\frac{i\sqrt{V_4}}{2w} \right)} \frac{w^{\frac{1}{2} -ip}\; {\rm HeunD}(\{ 1\};w)}{\sqrt{R+w\cos u}},
\end{equation}
where $p=V_3/\sqrt{V_4}, \kappa=\sqrt{k^2+V_0}$ and $\{1\}$ represents the set of five parameters of the double confluent Heun's function \cite{ronveaux, lay},
\begin{equation}
    \{1\} = \left\{-\frac{1}{4}+m^2-V_2 +p(i+p)+ \kappa\sqrt{V_4}, -2\kappa (p+i), -i\sqrt{V_4}, 2( 1-ip), -2i\kappa \right\}.
\end{equation}

The double confluent Heun's function $y(z)={\rm HeunD}(q, \alpha, \gamma, \delta, \varepsilon; z) $ satisfies the differential equation \cite{olver},
\begin{equation}
    z^2y^{''} + \left(\gamma+ \delta z+ \varepsilon z^2\right) y' + (\alpha z-q)y =0, 
\end{equation}
where its $\gamma$ also cannot be zero nor a negative integer.

The second solution is
\begin{equation}\label{psi2_potencial-U2-inversos}
    \psi^{(2)}_m(w, u, v) = d_m \exp{\left(imu -\frac{iv}{2} +i\kappa w -\frac{i\sqrt{V_4}}{2w} \right)} \frac{w^{\frac{1}{2} +ip}\; {\rm HeunD}(\{ 2\};w)}{\sqrt{R+w\cos u}},
\end{equation}
\begin{equation}
    \{2\} = \left\{-\frac{1}{4}+m^2-V_2 +p(-i+p)+ \kappa\sqrt{V_4}, V_1 -2\kappa (p-i) , i\sqrt{V_4}, 2( 1+ip), 2i\kappa   \right\}.
\end{equation}





The more familiar ordinary Bessel function is the solution of the ``radial-like" differential equation if one considers the simpler special case of \eqref{potencial-U2-inversos} with $V_1=V_3=V_4=0$. The non-normalized eigenfunctions are,
\begin{equation}
    \psi_m(w, u, v) =\frac{ \left[c_mJ_\ell\left( \kappa w\right) + d_m J_{-\ell}\left( \kappa w\right) \right]\exp\left(imu - \frac{iv}{2}\right)}{\sqrt{R+w\cos u}},
\end{equation}
where $\ell = \sqrt{m^2-V_2}$ is the ordinary Bessel function's order.
\subsection{Particle in a weak uniform magnetic field}
Immersed in a constant magnetic field $\mathbf{B} = B_0 \hat{k}$, a charged particle has a corresponding Schr\"{o}dinger equation that reads 
\begin{equation}
\begin{aligned}
    E\psi &= \frac{1}{2m^*}\left(-i\hbar\nabla - q\mathbf{A}\right)^2 \psi \\
          &= \frac{1}{2m^*}\left(-\hbar^2\nabla^2\psi + q^2 \mathbf{A}^2 \psi + 2i\hbar q\mathbf{A}\cdot\nabla\psi  + i\hbar q \psi \nabla \cdot \mathbf{A}\right),
\end{aligned}
\end{equation}
where $m^*$ is the particle's mass and $q$ is the electric charge. The choice of magnetic field implies $\mathbf{A} = (y\hat{i} - x\hat{j})B_0 / 2 = \hat{v}(R + w\cos u)B_0 / 2$, using the symmetric gauge, which in turn means $\nabla\cdot\mathbf{A} = 0$. Substituting the vector potential, we obtain, in the weak field approximation,

\begin{equation}\label{schrodinger com campo magnetico}
        -\hbar^2\nabla^2\psi  + i\hbar q B_0 \frac{\partial\psi}{\partial v} = 2m^*E\psi .
\end{equation}
It is straightforward to solve \eqref{schrodinger com campo magnetico}. Setting $k^2 = 2m^*E/\hbar^2$, one finds

\begin{equation}
    \psi_m(w, u, v) = B_m\frac{J_{m}(\kappa w) e^{i m u}\exp{\left(-i v/2\right)}}{\sqrt{R + w\cos u}},
\end{equation}
where $\kappa^2 = k^2 + q B_0/2\hbar$, i.e., a toroidal wavefunction with shifted energy eigenvalues.


\section{The Green's function and a plane-wave expansion}\label{sec:funcao-de-green}

In order to study the quantum scattering of a non-relativistic scalar particle, it is fundamental that we have a practical form of the Green function. As it is known \cite{arfken2005mathematical}, the Green function is the solution to 
\begin{equation}\label{eq de green}
    \nabla^2 G + k^2 G = -\delta(\mathbf{w} - \mathbf{w}').
\end{equation}

In toroidal-poloidal coordinates, we write the delta function as
\begin{equation}
    \delta(\mathbf{w} - \mathbf{w}') = \frac{\delta(w - w')\delta(u - u') \delta(v - v')}{w(R + w\cos u)} .
\end{equation}
An eigenfunction expansion can be achieved using
\begin{equation}
    \begin{aligned}
        \delta(\mathbf{w} - \mathbf{w}') &= \langle\psi(\mathbf{w'})|\psi(\mathbf{w})\rangle \\
                                         &= \sum_m \int_0^\infty dk k J_m(kw)J_m(kw')\frac{e^{i m (u - u')}e^{-i(v - v')/2}}{\sqrt{R + w\cos u}\sqrt{R + w'\cos u'}} \\
                                         &= \frac{\delta(w - w')}{w} \sum_m \frac{e^{i m (u - u')}e^{-i(v - v')/2}}{\sqrt{R + w\cos u}\sqrt{R + w'\cos u'}} 
    \end{aligned}
\end{equation}
A noteworthy feature of our basis is that it is not possible to construct an actual delta function of $v$ using only the two solutions we have, $e^{i v/2}$ and $e^{-i v/2}$. However, a simple calculation shows that our delta function does work as a filter regardless of this fact. We take inspiration from the eigenfunctions of our Schrödinger equation to write the bilinear form of our Green function \cite{hyperspherical,Cohl_2018,arfken2005mathematical} as
\begin{equation}\label{funcao de green} 
   G^{+}_0(\mathbf{w}|\mathbf{w}') = \frac{e^{-i( v- v')/2}}{\sqrt{R + w\cos u}\sqrt{R + w'\cos u'}}\sum_{m} g_m(w|w')e^{i m (u - u')}
\end{equation}
Then, substitution of \eqref{funcao de green} into the differential equation \eqref{eq de green} reduces it to
\begin{equation}\label{eq de green radial}
    \frac{d^2 g_m}{dw^2} + \frac{1}{w}\frac{d g_m}{dw} + \left(k^2 - \frac{m^2}{w^2}\right)g_m(w|w') = -\frac{\delta(w - w')}{w}.
\end{equation}
We can suppress the index $m$ and present two solutions,
\begin{align}
    g(w|w') =
    \begin{cases}
        g_1(w|w') = a_{m}(w')J_{m}(kw) + b_{m}(w')H_{m}^{(1)}(kw), \quad \text{if} \quad w < w' \\
        g_2(w|w') = c_{m}(w')J_{m}(kw) + d_{m}(w')H_{m}^{(1)}(kw), \quad \text{if} \quad w > w'
    \end{cases}
\end{align}
The boundary conditions are: the Green function must be continuous across the boundary $w = w'$, while its derivative is not. Additionally, the function is finite at $w = 0$, and we require that Sommerfeld's radiation condition be satisfied, that is, we want propagating waves from the origin to infinity, and not the opposite. The last two conditions are crucial, and they imply $b_{m}(w') = c_m(w') = 0$. The continuity condition at $w = w'$ ensures that 
\begin{equation}\label{cond1}
    a_{m}(w') J_{m}(kw')= d_{m}(w')H_{m}^{(1)}(kw'),
\end{equation}
while the discontinuity of the derivative of the radial Green function at $w = w'$ can be investigated by integrating equation \eqref{eq de green radial}. We get 
\begin{equation*}
   \lim_{w \rightarrow w'_+} \frac{d g_2}{dw} - \lim_{w \rightarrow w'_-}\frac{d g_1}{dw} = \frac{-1}{w'},
\end{equation*}
which translates to
\begin{equation}\label{cond2}
    -a_{m}(w') J'_{m}(kw') + d_{m}(w')H_{m}^{'(1)}(kw') = \frac{-1}{kw'}.
\end{equation}
We find that 
\begin{equation}
    a_m = \frac{i \pi}{2}H_{m}^{(1)}(kw'), \qquad d_m = \frac{i \pi}{2}J_{m}(kw') ,
\end{equation}
so using the traditional notation, we write the Green function of the Helmholtz equation in the toroidal/poloidal coordinate system
\begin{equation}\label{funcao-green-toroidal expansao} 
   G^{+}_0(\mathbf{w}|\mathbf{w}') = \frac{i \pi}{2} \frac{e^{i ( v'- v)/2} }{\sqrt{R + w\cos u}\sqrt{R + w'\cos u'}}\sum_{m} J_{m}(kw_<)H_{m}^{(1)}(kw_>)e^{i m (u - u')},
\end{equation}
with the notation $w_< = \min(w,w')$, $w_> = \max(w,w')$. To obtain a closed form of the Green function in these coordinates, notice that for a toroidally symmetric system with source at $\mathbf{w}' = 0$, equation \eqref{eq de green} reads
\begin{equation}
    \frac{\partial^2 G}{\partial w^2} + \left(\frac{1}{w} + \frac{\cos u}{R + w\cos u}\right)\frac{\partial G}{\partial w} + \frac{1}{(R + w\cos u)^2}\frac{\partial^2 G}{\partial  v^2} + k^2 G = -\frac{\delta(w - w')}{w}.
\end{equation}
Therefore, for $w \neq w'$, the incoming wave solution is

\begin{equation*}
    G(\mathbf{w}|\mathbf{w}') = B_0\frac{H_{0}^{(1)}(k|\mathbf{w} - \mathbf{w}'|)}{\sqrt{(R + w\cos u)e^{iv}}\sqrt{(R + w'\cos u')e^{iv'}}}.
\end{equation*}
Taking $w'\to 0$ and using
\begin{equation*}
    \lim_{\epsilon\to 0} \oint \frac{\partial G}{\partial w}\biggr\rvert_{w = \epsilon} dS_\epsilon = -1,
\end{equation*}
one calculates $B_0 = -1/16$. Thus we obtain,
\begin{equation}\label{funcao de green forma fechada}
    G(\mathbf{w}|\mathbf{w}') = -\frac{1}{16}\frac{H_{0}^{(1)}(k|\mathbf{w} - \mathbf{w}'|)}{\sqrt{(R + w\cos u)e^{iv}}\sqrt{(R + w'\cos u')e^{iv'}}}.
\end{equation}
This closed form of our Green function is symmetric, and we can use it to calculate an eigenfunction expansion for the plane wave.
\subsection{The Plane Wave}

In scattering problems, it is useful to use a plane wave as an impinging wave because of its simplicity and versatility. To calculate an eigenfunction expansion of a plane wave, notice that if we naively try
\begin{equation*}
    e^{i\mathbf{k}\cdot\mathbf{x}} = \sum_m B_m \frac{J_m(kw)e^{i m u}e^{-i v/2}}{\sqrt{R + w\cos u}}
\end{equation*}
we will end up facing seemingly innocent integrals that will result in expressions involving elliptic functions. Instead, contemplate taking the source to a position very far from our field point using equation \eqref{funcao-green-toroidal expansao}. The asymptotic behavior of our Hankel function is, with $z\to\infty$,
\begin{equation*}
    H_m^{(1)}(z) \approx \sqrt{\frac{2}{\pi z}}\exp{\left[i\left(z -\frac{m\pi}{2} - \frac{\pi}{4}\right) \right]}.
\end{equation*}
Since $w' \to \infty$, we also notice $|\mathbf{w} - \mathbf{w}'| \approx w' - \hat{\mathbf{n}}\cdot\mathbf{w}$. With this,
\begin{equation}\label{funcao-green-serie-aproximada} 
   G(\mathbf{w}|\mathbf{w}') \approx \frac{i \pi}{2} \frac{e^{i ( v'- v)/2} }{\sqrt{R + w\cos u}\sqrt{w'\cos u'}}\sqrt{\frac{2}{k\pi w'}}\frac{1}{\sqrt{2}}e^{i k w'}\sum_{m}(-i)^m J_{m}(kw)e^{i m (u - u')},
\end{equation}
Simultaneously, as we take $w' \to \infty$ to observe the closed form \eqref{funcao de green forma fechada} in its asymptotic behavior, we get

\begin{equation}
    G(\mathbf{w}|\mathbf{w}')  \approx -\frac{1}{16}\sqrt{\frac{1}{k\pi w'}}\frac{e^{ikw'}e^{-i\mathbf{k}\cdot\mathbf{w}}}{\sqrt{ w'\cos u' e^{iv'}}}\frac{1}{\sqrt{(R + w\cos u)e^{iv}}}.
\end{equation}
Comparing this with \eqref{funcao-green-serie-aproximada} results in
\begin{equation}\label{onda-plana-nova}
    e^{i\mathbf{k}\cdot\mathbf{w}} = -8\pi i\sum_{m} i^m J_{m}(kw) e^{im (u - u_k)} e^{-i v_k} .
\end{equation}
which has a surprisingly simple form. 

\section{Final remarks}\label{sec:conclusao}

We obtained, for the first time, the solution of the Helmholtz equation in Moon and Spencer's toroidal coordinates using a $\mathcal{R}$-separable solution, namely \eqref{solucao-matheus}. This represents a progress in the longstanding problem of analytically solving the Helmholtz equation in unorthodox coordinate systems. A similar approach was used in the toroidal--poloidal case, yielding a closed-form solution, \eqref{toroidal-wavefunctions}, in terms of Bessel's functions. This time, however, the problem is that the coordinate system only describes half of a torus, while covering the whole $\mathbb{R}^3$. Despite this, we went on to find the Green function and a plane-wave expansion. We also studied the quantum mechanics of a particle interacting with quite complicated potentials, \eqref{potencial-U1-nao-central} and \eqref{potencial-U2-inversos}, and presented exact solutions for the eigenfunctions in terms of bi-confluent and double-confluent Heun's functions. 

The inherent properties of this misleading geometry took long time to be understood, but we presented two solutions which may push this area further. We expect future research to continue from this point, bringing physical and engineering applications to this problem.

\begin{acknowledgments}
The authors gratefully acknowledge CNPq (grant numbers 140471/2022-7 and 309052/2023-8) for partial financial support.  M. E. Pereira would like to give special thanks to Alan Maioli for the insightful comments on the geometry of our coordinate system, H. S. Cohl for the inspiring conversations and to the whole Theoretical Physics Group at Universidade Federal Fluminense in Volta Redonda -- RJ for believing this work would be possible.

\end{acknowledgments}


\bibliography{apssamp}

\end{document}